\documentclass[preprint,aps,superscriptaddress,nofootinbib,tightenlines]{revtex4}

\usepackage{epsfig}

\def\OMIT#1{{}}

\newcommand{\dd}{{\cal Q}}
\newcommand{\dddd}{{\cal S}}

\newcommand{\nn}{\nonumber}
\newcommand{\tenb}{\bf{\overline{10}}}

\def\OMIT#1{{}}
\def\nc{{N_c}}

\def\parlam{^{\scriptstyle\lambda}}

\def\nc{{N_c}}

\def\yo2{{f_\pi^2}}
\def\llra{{\relbar\joinrel\longrightarrow}}
\def\mapright#1{{\smash{\mathop{\llra}\limits_{#1}}}}

\def\vev#1{{\langle #1\rangle}}

\begin{document}

\preprint{\vbox{
\hbox{UNH-04-05}
}}

\vphantom{}

\title{Stable Pentaquarks from Strange Chiral Multiplets}

\author{\bf Silas R.~Beane}
\affiliation{Department of Physics, University of New Hampshire,
Durham, NH 03824-3568.}
\affiliation{Jefferson Laboratory, 12000 Jefferson Avenue, 
Newport News, VA 23606.}

\vphantom{}
\vskip 0.5cm
\begin{abstract} 
\vskip 0.5cm
\noindent 
The assumption of strong diquark correlations in the QCD spectrum
suggests flavor multiplets of hadrons that are degenerate in the
chiral limit. Generally it would be unnatural for there to be
degeneracy in the hadron spectrum that is not protected by a QCD
symmetry. Here we show ---for pentaquarks constructed from diquarks---
that these degeneracies can be naturally protected by the full chiral
symmetry of QCD. The resulting chiral multiplet structure recovers the
ideally-mixed pentaquark mass spectrum of the diquark model, and
interestingly, requires that the axial couplings of the pentaquarks to
states outside the degenerate multiplets vanish in the chiral
limit. This result suggests that if these hadrons exist, they are
stable in the chiral limit and therefore have widths that scale as the
fourth power of the kaon mass over the chiral symmetry breaking
scale. Natural-size widths are of order a few MeV.
\end{abstract}

\maketitle

\vfill\eject



\section{Introduction}
\label{sec:intro}

\noindent Partly motivated by color-superconducting phases of QCD at
asymptotic baryon densities~\cite{Alford:1998mk,Son:1998uk,Krishna}, Jaffe and
Wilczek (J-W) have recently argued~\cite{Jaffe:2003sg} that there are
strong diquark correlations in low-energy QCD and that these
correlations may show up as novel hadronic states that do not readily
fit into the standard constituent-quark picture. The recent
highly-controversial observation~\cite{Nakano:2003qx,Barmin:2003vv,Stepanyan:2003qr,Barth:ja,
Asratyan:2003cb,Alt:2003vb,Kubarovsky:2003fi,Airapetian:2003ri,Aleev:2004sa,Chekanov:2004kn}
of exotic ``pentaquark'' states has provided renewed interest in the
idea that diquark correlations are important for hadron
spectroscopy~\cite{Jaffe:2003sg,Karliner:2003dt,Jaffe:2003ci,Jaffe:2004zg,Stewart:2004pd,Zhang:2004xt}.
While the pentaquark is now recognized by the particle-data group~\cite{pdg},
there are several null results from high-energy experiments~\cite{Bai:2004gk,Knopfle:2004tu,Pinkenburg:2004ux,Antipov:2004jz}.
Experiments with an expected ten-fold increase in statistics using deuterium and hydrogen targets 
are currently being analyzed from CLAS at JLab. It is hoped that high statistics will determine whether 
or not pentaquarks exist~\footnote{A recent critical review of the current state-of-affairs is given in Ref.~\cite{Hicks}.}.
Lattice QCD simulations have been performed~\cite{Csikor:2003ng,Sasaki:2003gi,Chiu:2004gg,Mathur:2004jr,Csikor:2004us}, 
however, to date, the results are inconclusive.

A puzzling feature that any theoretical description of pentaquarks
must face is their extremely narrow widths; measurements of the widths
are limited by the current experimental
resolution~\cite{Nakano:2003qx,Barmin:2003vv,Stepanyan:2003qr,Barth:ja,
Asratyan:2003cb,Alt:2003vb,Kubarovsky:2003fi,Airapetian:2003ri,Aleev:2004sa,Chekanov:2004kn}.
An important point regarding the expected size of the pentaquark
widths has been made in Ref.~\cite{Jenkins:2004tm}. We will repeat the
argument here.  In estimating the $\Theta^+\,N\,K$ axial coupling, one
should not compare to, for instance, the $\Delta\,N\,\pi$ axial
coupling which is a number of order one, but rather to the
$N^*\,N\,\pi$ axial coupling, where $N^*$ is an excited baryon, which
is generically a number much less than one~\footnote{For a tabulation
of these axial couplings, see Ref.~\cite{Beane:2002ud}.}.  One useful
way to think about this empirical fact is offered by the large-$\nc$
approximation~\cite{largeNearly}, where $\Theta$ and $N$ appear in
different irreducible representations of the spin-flavor symmetry
group, as do $N$ and $N^*$, while $N$ and $\Delta$ are in the same
irreducible representation.  This symmetry structure implies that the
$\Theta^+\,N\,K$ and $N^*\,N\,\pi$ axial couplings are suppressed in
the large-$\nc$ counting as compared to the $\Delta\,N\,\pi$ axial
coupling~\cite{Jenkins:2004tm}.  This paper will address the issue of the expected size of
the pentaquark widths using symmetry arguments that, while unrelated to
the large-$\nc$ approximation, are similar in spirit.

The diquark picture leads one to expect degeneracies in the chiral
limit between $SU(3)$ multiplets that are not obviously protected by
any QCD symmetry. (Here we will consider light quarks only, $q\sim
u,d,s$.)  For instance, for tetraquarks ($qq\overline q\overline q$)
one expects a degenerate nonet, ${\bf 1}\oplus{\bf 8}$, and for
pentaquarks ($qqqq\overline q$) one expects a degenerate ${\bf
8}\oplus{\bf\overline{10}}$ of positive parity as well as a degenerate
${\bf 1}\oplus{\bf 8}$ of negative parity. These multiplets
are expected to appear with a variety of spin content. The
expectations of degeneracy are based on the assumption that
flavor-dependent interactions between the quarks and the antiquarks
are absent. Of course, it would be unnatural for there to be
degeneracy in the hadron spectrum that is not protected by a symmetry
of QCD.  For instance, in the large-$\nc$ limit~\cite{largeNearly} one
naturally expects singlet-adjoint degeneracy in the meson spectrum;
that is because the anomaly is suppressed in this limit and the flavor
symmetries enhance from $SU(3)$ to $U(3)$. Here we will see that the
$SU(3)_R\otimes SU(3)_L$ chiral symmetry of QCD can require
singlet-adjoint and octet-antidecuplet degeneracy in the low-energy
theory.  

It may seem perplexing that chiral symmetry can protect a degeneracy
in the low-energy theory as the hadronic Hamiltonian clearly has a
contribution that breaks chiral symmetry, even in the chiral
limit. However, by working in collinear Lorentz frames, one can see that
there are no non-vanishing matrix elements of the symmetry-breaking
part of the Hamiltonian between hadrons that are in irreducible
representations of the chiral group. Therefore the states within
irreducible representations can remain degenerate as a consequence of
chiral symmetry even though the Hamiltonian has a large
symmetry-breaking piece.  Now the remarkable thing is that states in
different chiral multiplets cannot communicate by Goldstone-boson
emission and absorption. Therefore, in what we will refer to as ``the
natural J-W scenario'', with chiral symmetry protecting the degeneracy
between flavor multiplets, the chiral-limit axial couplings of the
tetraquarks and the pentaquarks to the ordinary mesons and baryons
must vanish. In particular, this implies that the tetraquarks and
pentaquarks built from diquarks are stable in the chiral limit. Their
decay widths to the ordinary mesons and baryons generally scale as
$M_q^2$, where $M_q$ is the quark mass matrix, and are therefore
suppressed.

This paper is organized as follows. In Sec.~\ref{sec:diquarkfields} we
construct the pentaquark interpolating fields from diquarks and write
down the leading-order (LO) operators in chiral perturbation theory ($\chi$PT)
that govern the pentaquark axial transitions. In
Sec.~\ref{sec:chiralmultssu3} we introduce technology which allows one
to extract consequences of $SU(3)_R\otimes SU(3)_L$ for hadrons and we
consider possible representations filled out by the pentaquarks.  We
then focus on the diquark scenario. In Sec.~\ref{sec:AW} we recover
the pentaquark axial couplings obtained from the chiral algebra in the
diquark scenario using generalized Adler-Weisberger sum rules. We conclude in
Sec.~\ref{sec:conc}.

\section{Fields and Operators}
\label{sec:diquarkfields}

\subsection{Diquark Field}

\noindent The basic assumption of the J-W picture is that
quarks correlate strongly in the channel which is antisymmetric in color, spin, and
flavor. For light quarks the resulting bosonic diquark, $\dd$, is
a color and flavor-$SU(3)$ antitriplet with $J^{P}=0^{+}$. That is,
\begin{eqnarray}
\dd^{a \alpha}\ = \ \epsilon^{abc} \epsilon^{\alpha\beta\gamma}\ q_{b\beta}\ q_{c\gamma} \ ,
\label{eq:diquarkQ}
\end{eqnarray}  
where roman indices are fundamental flavor and greek indices are fundamental color.
This diquark object is argued by J-W to be an important degree of freedom in low-energy QCD.
We will now construct pentaquark interpolating fields out of diquarks.

\subsection{Pentaquark Fields}

\noindent As diquarks are flavor-$SU(3)$ antitriplets, the only way to
make an exotic pentaquark out of two diquarks and an antiquark is to
combine the diquarks symmetrically in flavor, $(\overline{\bf 3}
\otimes \overline{\bf 3})_{\cal S} = \overline{\bf 6}$, and then
couple the antiquark.  The flavor content of the resulting lowest-lying
$qqqq\overline q$ states is then a degenerate $\overline{\bf 6}\otimes
\overline{\bf 3}= {\bf 8}\oplus \overline{\bf 10}$ with
spin-parity $\textstyle{1\over2}^{+}$. Neglecting the
color indices, the $\overline{\bf 6}$ may be represented by the
symmetric tensor~\cite{Oh:2004gz},
\begin{eqnarray}
\dddd^{ab} \equiv\ {1\over{2\sqrt{2}}}\ \left(\ \dd^{a} {\dd}^{b} \ +\ \dd^{b} {\dd}^{a} \ \right)\ .
\label{eq:Sabdefined}
\end{eqnarray}  
We can then contract the $\overline{\bf 6}$ with the $\overline{\bf 3}$ antiquark to give the interpolator, 
${\cal T}^{abc}$, for the ${\bf 8}\oplus \overline{\bf 10}$ pentaquarks:  
\begin{eqnarray}
{\cal T}^{abc}\ =\ {1\over\sqrt{2}}\dddd^{ab}\ \overline{q}^c \ = 
\ P^{abc}\ +\ {1\over\sqrt{6}}\left(\ \epsilon^{dbc}\ {{\cal O}}^a_d \ +\ \epsilon^{dac}\ {{\cal O}}^b_d \ \right) \ ,
\label{eq:pentaquarkinterp}
\end{eqnarray}  
where
\begin{eqnarray}
P^{abc}\ &=& \  {1\over{3\sqrt{2}}}\ \left(\ \dddd^{ab}\ \overline{q}^c\ +\ \dddd^{ac}\ \overline{q}^b\ +\ \dddd^{bc}\ \overline{q}^a 
\ \right)\ ;
\nn \\
{{\cal O}}^a_b \ &=& \ {1\over{\sqrt{3}}}\ \epsilon_{bcd}\ \dddd^{ac}\ \overline{q}^d \ .
\label{eq:pentaquarkinterpquarks}
\end{eqnarray}  
In terms of the hadronic description we have the antidecuplet
\begin{eqnarray}
P^{333} &=& \Theta^{+}  \nn \\
P^{133} = \frac{1}{\sqrt{3}} \, N^0_{\tenb}  & , & 
P^{233} = \frac{1}{\sqrt{3}} \, N^+_{\tenb} \nn \\
P^{113} = \frac{1}{\sqrt{3}} \, \Sigma^{-}_{\tenb}\quad ,\quad 
P^{123} &=& \frac{1}{\sqrt{6}}\Sigma^{0}_{\tenb}\quad , \quad
P^{223} = \frac{1}{\sqrt{3}} \, \Sigma^{+}_{\tenb}   \nn \\
P^{111} =  \Xi^{--}_{\tenb} \qquad ,\qquad
P^{112} = \frac{1}{\sqrt{3}}  \, \Xi_{\tenb}^{-}  &\quad  ,\quad & 
P^{122} = \frac{1}{\sqrt{3}} \, \Xi_{\tenb}^{0} \qquad ,\qquad
P^{222} =  \Xi_{\tenb}^{+}\ ,
\label{eq:pentacomponents}
\end{eqnarray}  
and the octet
\begin{eqnarray}
{\hat {{\cal O}}} & = & \left(\matrix{{1\over\sqrt{6}}\Lambda_{\cal O} + {1\over\sqrt{2}}\Sigma^0_{\cal O}
& \Sigma_{\cal O}^+ & p_{\cal O}\cr
\Sigma_{\cal O}^- & {1\over\sqrt{6}}\Lambda_{\cal O} - {1\over\sqrt{2}}\Sigma_{\cal O}^0 & n_{\cal O}\cr
\Xi_{\cal O}^- & \Xi_{\cal O}^0 & -{2\over\sqrt{6}}\Lambda_{\cal O}}\right)
\ \ \ .
\label{eq:pentoctet}
\end{eqnarray}
A priori, in QCD, we would expect that the $P$-${\cal O}$ mass splitting in the chiral limit is of order $\Lambda_{QCD}$.
In the diquark picture of J-W the octet and antidecuplet are degenerate
in the chiral limit. This implies that the low-energy effective field theory should be formulated
using the field ${\cal T}^{abc}$ rather the $P$ and ${\cal O}$ fields separately; i.e. there should
exist a symmetry which places $P$ and ${\cal O}$ in a single $18$-dimensional multiplet.
We will see below that chiral symmetry can fill this role.

\subsection{Pentaquark Effective Lagrangian}

\noindent  The pentaquark octet is described by a
two-index tensor, ${{\cal O}}_a^b$, and transforms as
\begin{eqnarray}
{\cal O}_a^b\ \rightarrow \ (U)^{a'}_a (U^\dagger)^b_{b'} {\cal O}_{a'}^{b'} \ 
\label{eq:pentaoctettrans}
\end{eqnarray}
with respect to the diagonal flavor-$SU(3)$ subgroup of $SU(3)_R\otimes SU(3)_L$. 
The pentaquark octet matrix is given in Eq.~(\ref{eq:pentoctet}).
The baryon octet is described by a two-index tensor, ${B}_a^b$, which 
transforms in the same way as ${{\cal O}}_a^b$ with respect to flavor and is given by
\begin{eqnarray}
{\hat {{B}}} & = & \left(\matrix{{1\over\sqrt{6}}\Lambda + {1\over\sqrt{2}}\Sigma^0
& \Sigma^+ & p\cr
\Sigma^- & {1\over\sqrt{6}}\Lambda - {1\over\sqrt{2}}\Sigma^0 & n\cr
\Xi^- & \Xi^0 & -{2\over\sqrt{6}}\Lambda}\right)
\ \ \ .
\label{eq:baroctet}
\end{eqnarray}
The antidecuplet pentaquarks are described by a three-index completely-symmetric tensor, $P^{abc}$, and transform as
\begin{eqnarray}
P^{abc}\ \rightarrow \ (U^\dagger)_{a'}^a (U^\dagger)_{b'}^b (U^\dagger)_{c'}^c
P^{a'b'c'}
\label{eq:antidecuplettrans}
\end{eqnarray}
with respect to flavor-$SU(3)$. The components of the antidecuplet tensor are given in Eq.~(\ref{eq:pentacomponents}).

At LO in the three-flavor chiral expansion, the relevant
axial matrix elements are parametrized by the pentaquark self-couplings, 
${\cal D}_{\cal O}$, ${\cal F}_{\cal O}$, ${\cal H}_P$, ${\cal C}_{P{\cal O}}$ and 
the couplings of the pentaquarks to the baryons, ${\cal C}_{PB}$, ${\cal D}_{{\cal O}B}$ and ${\cal F}_{{\cal O}B}$.
Assuming $J^p=\textstyle{1\over2}^{+}$ pentaquarks, the LO $\chi$PT lagrangian~\cite{Jenkins:1991es,Ko:2003xx,Mehen:2004dy,Mohta:2004yf} is
\begin{eqnarray}
{\cal L} &=&  2\, {\cal D}_{\cal O}\, \bar{{\cal O}}\, S^\mu\, \left\{{\cal A}_\mu,{\cal O}\right\} \ +\
2\, {\cal F}_{\cal O}\,  \bar{{\cal O}}\, S^\mu\, \left[{\cal A}_\mu,{\cal O}\right]  \ +\ 
2\, {\cal H}_{P}\, \bar{P}\, (S \cdot {\cal A})\, P  \nn \\
& &\quad\quad \ +\ 2\, {\cal C}_{P{\cal O}}\, \lbrack\ \bar{P}\, (S \cdot {\cal A})\, {\cal O} + h.c.\ \rbrack\ +\  
2\, {\cal C}_{PB}\, \lbrack\ \bar{P}\, (S \cdot {\cal A})\, B + h.c.\ \rbrack  \nn \\
& &\quad\quad\quad \ +\ 2\, {\cal D}_{{\cal O}B}\, \lbrack\ \bar{B}\, S^\mu\, \left\{{\cal A}_\mu,{\cal O}\right\} + h.c.\ \rbrack \ +\
2\, {\cal F}_{{\cal O}B}\,  \lbrack\ \bar{B}\, S^\mu\, \left[{\cal A}_\mu,{\cal O}\right] + h.c.\ \rbrack \ ,
\label{eq:chiptlagsu3pentas}
\end{eqnarray}
where ${\cal A}_\mu$ is the axial-vector field that contains the Goldstone bosons,
and $S_\mu$ is the usual spin operator~\footnote{See Refs.~\cite{Jenkins:1991es,Ko:2003xx,Mehen:2004dy,Mohta:2004yf} for index contractions
and other details.}. This lagrangian is defined in the chiral limit; when the quark masses are turned on
mixing occurs and the diagonalization of the mass matrix shifts the axial couplings. 

\section{$SU(3)\otimes SU(3)$ Representations}
\label{sec:chiralmultssu3}

\subsection{The Charge Algebra}

\noindent In this section we will develop minimal technology that will allow
us to extract consequences of chiral symmetry in low-energy
QCD. Consider three-flavor QCD in the chiral limit. Weinberg and others have shown
that by working in Lorentz frames in which all momenta are
collinear, one may use the full $SU(3)_R\otimes SU(3)_L$ symmetry of QCD to
classify hadrons~\cite{Harari:1966yq,lipkin,gerstein,Harari:1966jz,Gilman,Weinberga,Weinbergb,strong,Beane:2002td}. 
As helicity is conserved in collinear frames, the chiral classification is for each helicity, as we will see.
The presence of the full chiral-symmetry group in the low-energy theory is related to
the special asymptotic behavior of certain Goldstone-boson-hadron scattering
amplitudes. Hence, not surprisingly, consequences of chiral symmetry
may also be extracted from the study of generalized Adler-Weisberger
sum rules~\cite{AW,AW2}, whose validity follows from the special asymptotic constraints.
We express the algebra of $SU(3)_R\otimes SU(3)_L$ via~\footnote{Here we
use greek indices for adjoint flavor.}
\begin{equation}
\lbrack {{{X}}_{\alpha}\parlam},\, {{{X}}_{\beta}\parlam}\rbrack
=i f_{\alpha\beta\gamma} {T}_{\gamma}\quad ,\quad
\lbrack {T}_{\alpha},\,{{{{X}}_{\beta}\parlam}}\rbrack= i f _{\alpha\beta\gamma}{{{{X}}_{\gamma}\parlam}} \quad ,\quad
\lbrack {T}_{\alpha},\,{T}_{\beta}\rbrack=i f_{\alpha\beta\gamma}{T}_{\gamma}
\label{eq:basicchirasymmsu3}
\end{equation}
where $T_\alpha=\lambda_\alpha/2$. The object ${{X}}_{\alpha}\parlam$ acts as an axial-vector current or charge in the low-energy theory; its
matrix elements between hadron states mediate Goldstone boson emission and absorption. The Feynman amplitude
for a Goldstone-boson transition between hadrons ${\cal A}$ and ${\cal B}$ of helicity $\lambda$
is 
\begin{eqnarray}
M( {\cal A}\rightarrow {\cal B}\, \pi_\alpha\,;\, \lambda)\ =\ {1\over{F_\pi}}(\, M_{\cal A}^2\ -\ M_{\cal B}^2\, )
\langle\,  {\cal A}\, ,\, \lambda\, | \, X^{\,\lambda}_\alpha\, |\, {\cal B}\, ,\, \lambda\, \rangle \ ,
\label{eq:Feynmanamp}
\end{eqnarray}
where $F_\pi=93~{\rm MeV}$.
The LO pentaquark axial-vector current may be found from the $\chi$PT lagrangian of Eq.~(\ref{eq:chiptlagsu3pentas})
and is given by
\begin{eqnarray}
(X^{\,\uparrow}_\alpha )_{LO}&=& \ {\cal D}_{\cal O}\  {\cal O}^\dagger_\uparrow \ \lbrace {T_\alpha},{\cal O}^{{}}_\uparrow \rbrace  +
{\cal F}_{\cal O}\  {\cal O}^\dagger_\uparrow\  \lbrack
{T_\alpha},{\cal O}^{{}}_\uparrow\ \rbrack\ +\ {\cal H}_P\ 
P^\dagger_\uparrow\ {T_\alpha}\ P^{{}}_\uparrow  \nn \\
& &\quad\quad \ +\ {\cal C}_{P{\cal O}}\ \lbrack\ {\cal O}^\dagger_\uparrow\  {T_\alpha}\ P^{{}}_\uparrow  +  h.c. \rbrack\ +\  
{\cal C}_{PB}\ \lbrack\ B^\dagger_\uparrow\  {T_\alpha}\ P^{{}}_\uparrow  +  h.c. \rbrack  \nn \\
& & \quad\quad\quad \ +\ {\cal D}_{{\cal O}B}\ \lbrack\ B^\dagger_\uparrow \ \lbrace {T_\alpha},{\cal O}^{{}}_\uparrow \rbrace 
+  h.c. \rbrack  +
{\cal F}_{{\cal O}B}\ \lbrack\ B^\dagger_\uparrow\  \lbrack {T_\alpha},{\cal O}^{{}}_\uparrow\ \rbrack +  h.c. \rbrack
\ ,
\label{eq:chiptcurrentsu3pentas}
\end{eqnarray}
where we have projected out the $\lambda=1/2$ ($\uparrow$) current. The $\lambda=-1/2$ ($\downarrow$) current is given by
$X^{\,\downarrow}_\alpha =-X^{\,\uparrow}_\alpha $. 
As an example, the helicity-$\uparrow$ transition $\Theta^+\rightarrow K^+ n$ is mediated by the axial-vector matrix element
\begin{eqnarray}
\langle\, \Theta^+\, \uparrow\, | \, X^{\,\uparrow}_4\, -\, i\, X^{\,\uparrow}_5\, |\, n\, \uparrow\, \rangle \, =\, -{\cal C}_{PB} \ .
\end{eqnarray}

The constraints imposed by the algebra of Eq.~(\ref{eq:basicchirasymmsu3}) on the current of Eq.~(\ref{eq:chiptcurrentsu3pentas}) are determined
by how the pentaquark states are placed in representations of the chiral group.
We should therefore consider the allowed chiral representations.
Now the quarks transform as $(\bf{1},\bf{3})$ and $(\bf{3},\bf{1})$, with respect to
$(SU(3)_R,SU(3)_L)$. In a helicity-conserving frame, helicity and chirality are the same. Therefore, if
the helicity-$\lambda$ component of a hadron is in the 
$({\bf R_1},{\bf R_2})$ representation of
$SU(3)_R\otimes SU(3)_L$ where ${\bf R_{1,2}}$ are flavor-$SU(3)$ representations, then 
the $-\lambda$ component of the hadron is in the 
$({\bf R_2},{\bf R_1})$ irreducible representation.
We will now review the allowed chiral representation for the baryons.
The flavor-$SU(3)$ decomposition of a baryon interpolating operator is
\begin{eqnarray}
q\,q\,q\,\sim\, {\bf 1}\oplus{\bf 8}_2\oplus{\bf 10} \ , 
\label{eq:lightinterpolator3}
\end{eqnarray}
where the subscript denotes multiplicity.  
Therefore, generally, we expect the baryons to transform
as linear combinations of $(\bf{\overline{3}},\bf{3})$, $(\bf{{3}},\bf{\overline{3}})$
$(\bf{3},\bf{6})$, $(\bf{6},\bf{3})$, $(\bf{8},\bf{1})$, $(\bf{1},\bf{8})$,
$(\bf{10},\bf{1})$, $(\bf{1},\bf{10})$ and $(\bf{1},\bf{1})$ irreducible representations
as these are the only representations of $SU(3)_R\otimes SU(3)_L$ that contain the
flavor multiplets of the baryon interpolator and no others.

The $SU(3)$ decomposition of a pentaquark interpolating operator is
\begin{eqnarray}
q\,q\,q\,q\,\bar{q}\sim\, {\bf 1}_3\oplus{\bf 8}_8\oplus{\bf 10}_4\oplus{\bf\overline{10}}_2\oplus{\bf 27}_3\oplus{\bf 35} \ .
\label{eq:pentainterpolator}
\end{eqnarray}
In addition to the chiral representations allowed for the baryons, the
pentaquarks may also be in linear combinations of $(\bf{\overline{10}},\bf{1})$,
$(\bf{1},\bf{\overline{10}})$,
$(\bf{\overline{3}},\bf{\overline{6}})$,
$(\bf{\overline{6}},\bf{\overline{3}})$, $(\bf{27},\bf{1})$,
$(\bf{1},\bf{27})$, $(\bf{35},\bf{1})$, $(\bf{1},\bf{35})$,
$(\bf{8},\bf{8})$, $(\bf{8},\bf{10})$, $(\bf{10},\bf{8})$,
$(\bf{8},\bf{\overline{10}})$, $(\bf{\overline{10}},\bf{8})$,
$(\bf{6},\bf{\overline{6}})$ and $(\bf{\overline{6}},\bf{6})$
irreducible representations of $SU(3)_R\otimes SU(3)_L$.

In Ref.~\cite{Beane:2002td} it was conjectured that the chiral multiplet filled out by a given
hadron is the minimal representation that contains only the flavor multiplets of the interpolator
for that hadron. Evidence was provided in two-flavor QCD for light and heavy-light systems
that this is indeed the case.
Given the vast multiplicity of the pentaquark interpolator of Eq.~(\ref{eq:pentainterpolator}) , it is not clear that the conjecture
of Ref.~\cite{Beane:2002td} has anything useful to say about it.
By constrast, in the diquark picture, the $SU(3)$ decomposition of a pentaquark interpolating operator is
\begin{eqnarray}
\dd \,\dd \,{\overline{q}}\,\sim\, ({\bf 1}\oplus{\bf 8})_-\oplus({\bf 8}\oplus{\bf\overline{10}})_+ \ ,
\label{eq:pentainterpolatordiquark}
\end{eqnarray}
where the subscripts denote parity. Therefore, here we expect the pentaquarks to transform
as linear combinations of $(\bf{\overline{3}},\bf{3})$, $(\bf{{3}},\bf{\overline{3}})$
$(\bf{\overline{3}},\bf{\overline{6}})$, $(\bf{\overline{6}},\bf{\overline{3}})$, $(\bf{8},\bf{1})$, $(\bf{1},\bf{8})$,
$(\bf{\overline{10}},\bf{1})$, $(\bf{1},\bf{\overline{10}})$ and $(\bf{1},\bf{1})$ irreducible representations.
A degenerate octet and antidecuplet naturally fall into $(\bf{\overline{3}},\bf{\overline{6}})$ and  
$(\bf{\overline{6}},\bf{\overline{3}})$ irreducible representations and a
degenerate singlet and octet naturally fall into $(\bf{\overline{3}},\bf{3})$ and  $(\bf{{3}},\bf{\overline{3}})$
irreducible representations. The diquark interpretation of pentaquarks is therefore consistent with
the conjecture of Ref.~\cite{Beane:2002td}. Below we will place the $\bf{8}\oplus\bf{\overline{10}}$ pentaquarks
in the $(\bf{\overline{3}},\bf{\overline{6}})$ representation of $SU(3)_R\otimes SU(3)_L$.

\subsection{The Mass-Squared Matrix}

\noindent In helicity-conserving frames, spontaneous chiral symmetry breaking appears in the hadronic
Hamiltonian as an operator that transforms non-trivially with respect to $SU(3)_R\otimes SU(3)_L$.
We will assume that the hadronic mass-squared matrix~\footnote{We work with the mass-squared matrix as that
is the relevant quantity in helicity conserving frames. An easy way to see this is to take the infinite-momentum
limit of the relativistic energy dispersion relation.} can be decomposed as:
\begin{eqnarray}
{{\hat M}^{2}}\ =\ {{\hat M}^{2}_{\bf{1}}}\ +\ {{\hat M}^{2}_{\bf{3}\bf{3}}} \ \ ,
\label{eq:basicMsquaredsu3} 
\end{eqnarray}
where ${{\hat M}^{2}_{\bf{1}}}$ transforms as a singlet,
$(\bf{1},\bf{1})$, and ${{\hat M}^{2}_{\bf{3}\bf{3}}}$ transforms as
$(\bf{\bar{3}},\bf{3})\oplus(\bf{3},\bf{\bar{3}})$. This last assumption is not crucial; it is sufficient
for our purposes that ${{\hat M}^{2}}$ contain a piece that transforms non-trivially with
respect to the chiral group in the sense
that $\lbrack {{{X}}_{\alpha}\parlam}, {{\hat M}^{2}}\rbrack\neq 0$ since in the absence of this piece 
there would be no spontaneous chiral-symmetry breaking in the low-energy theory.
This then allows us to prove several useful lemmas~\cite{Harari:1966yq,lipkin,gerstein,Harari:1966jz,Gilman,Weinberga,Weinbergb,strong,Beane:2002td}. 
Define $\lbrack {{{X}}_{\alpha}\parlam}, {{\hat M}^{2}} \rbrack\equiv {{\hat {\cal M}}_\alpha^{2}}$.
Taking the matrix element of this relation between hadron states ${\cal A}$ and ${\cal B}$ with helicity-$\lambda$ gives
\begin{eqnarray}
(\,M_{\cal B}^2\ -\ M_{\cal A}^2\,)
\langle\,  {\cal A}\, ,\, \lambda\, | \, X^{\,\lambda}_\alpha\, |\, {\cal B}\, ,\, \lambda\, \rangle \ =\ 
\langle\,  {\cal A}\, ,\, \lambda\, | \, {{\hat {\cal M}}_\alpha^{2}} \, |\, {\cal B}\, ,\, \lambda\, \rangle \ .
\label{eq:theorem}
\end{eqnarray}

\vskip0.2in

\noindent LEMMA 1: Say ${\cal A}$ and ${\cal B}$ are in an irreducible
representation. Then the right-hand side of Eq.~(\ref{eq:theorem})
vanishes as ${\cal A}$ and ${\cal B}$ overlap only through the singlet
part of the mass-squared matrix, ${{\hat M}^{2}_{\bf{1}}}$. As  $X_\alpha$ acts as a symmetry generator,
$\langle\,{\cal A}\, | \, X_\alpha\, |\, {\cal B}\,\rangle$ is nonvanishing for ${\cal A}$ and
${\cal B}$ in the same irreducible representation. It follows that {\it if hadrons
${\cal A}$ and ${\cal B}$ are in an irreducible representation, they must
be degenerate}.

\vskip0.2in

\noindent LEMMA 2:  Say ${\cal A}$ and ${\cal B}$ are in a reducible representation. 
A representation is reducible if any two irreducible representations that make up the reducible multiplet
have a non-vanishing overlap with the symmetry breaking part of the mass-squared matrix, ${{\hat M}^{2}_{\bf{3}\bf{3}}}$.
Hence the right hand side of Eq.~(\ref{eq:theorem}) will be nonvanishing, as will the
matrix element for Goldstone boson transitions via Eq.~(\ref{eq:Feynmanamp}).
If ${\cal A}$ and ${\cal B}$ are in different chiral representations, 
the right hand side of Eq.~(\ref{eq:theorem}) vanishes. 
Therefore, {\it hadrons in different chiral representations do not communicate by Goldstone boson emission and
absorption}. 

\vskip0.1in

These lemmas will be crucial in what follows.

\subsection{Explicit Breaking Effects}

\noindent We will now consider quark-mass corrections. In QCD the quark mass matrix, ${\hat M_q}$, transforms as 
${\hat M_q} \ \rightarrow\  L\ {\hat M_q} \  R^\dagger $ with respect to
$SU(3)_R\otimes SU(3)_L$. The quark mass matrix is 
${\hat M}_q  = {\rm diag}(m,m,m_s)$ where $m$ is the common light-quark mass, and
$m_s$ is the strange quark mass. In order to properly introduce explicit
breaking effects in the helicity conserving theory, it is convenient to introduce in
addition a spurion, $\vev{c}$, with dimensions of mass which
transforms like the quark mass matrix and acts like an effective condensate
parameter in the low-energy theory. Inclusion of explicit breaking effects
is then achieved by allowing invariant operators with insertions of
$\vev{c}\,M_q$, which transforms as
\begin{eqnarray}
&&(\vev{c}\,M_q)_L \ \rightarrow\  L\ (\vev{c}\,M_q)_L\  L^\dagger
\ \ ,\ \ 
(\vev{c}\,M_q)_R \ \rightarrow\  R\ (\vev{c}\,M_q)_R\  R^\dagger
\ \ .
\label{eq:Mqsigtransfsu3}
\end{eqnarray}
This transformation rule amounts to assuming that explicit breaking effects
are purely octet with respect to flavor-$SU(3)$. We should therefore recover the Gell-Mann-Okubo formula
for the baryon octet and the equal-spacing relations for the baryon decuplet.
In what follows we define ${\cal X}\equiv\langle s\rangle\,M_q$. Our final
form for the hadronic mass-squared matrix is
\begin{eqnarray}
{{\hat M}^{2}}\ =\ {{\hat M}^{2}_{\bf{1}}}\ +\ {{\hat M}^{2}_{\bf{3}\bf{3}}} 
\ +\ {{\hat M}^{2}_{q\bf{8}}} \ \ ,
\label{eq:basicMsquaredsu3full} 
\end{eqnarray}
where ${{\hat M}^{2}_{q\bf{8}}}$ is from the quark masses and transforms as
$(\bf{1},\bf{8})\oplus(\bf{8},\bf{1})$. The lemmas proved above continue
to hold away from the chiral limit but now ${{\hat {\cal M}}_\alpha^{2}}$
has a component that transforms as $(\bf{1},\bf{8})\oplus(\bf{8},\bf{1})$
and which gives rise to new mixing between chiral multiplets. We will
see an example of this mixing below.

\subsection{Pentaquark Octet and Antidecuplet in a $(\overline{{\bf 3}},\overline{{\bf 6}})$}

\subsubsection{Field Content and the Mass-Squared Matrix}

\noindent 
In this section we will put the helicity-$\uparrow$ ${\cal O}$ and 
$P$ in the $(\overline{{\bf 3}},\overline{{\bf 6}})$ irreducible representation
of $SU(3)_R\otimes SU(3)_L$. We introduce the three-index tensor, ${\cal T}^{a,bc}$, which transforms as
\begin{eqnarray}
&& {\cal T}^{a,bc}\  \rightarrow\ 
( R^\dagger )^a_{a'} ( L^\dagger )^b_{b'} ( L^\dagger )^c_{c'}
{\cal T}^{a' ,b'c'}
\ \ \ .
\label{eq:TantiRLLsu3}
\end{eqnarray}
The helicity-$\downarrow$ ${\cal O}$ and $P$ may be placed 
in an analogous tensor transforming in the $(\overline{{\bf 6}},\overline{{\bf 3}})$ representation.
The tensor ${\cal T}$ must be symmetric under the interchange of the two left handed
indices, while there is no symmetry condition for the interchange of right
and left handed indices. In terms of tensors transforming as 
an $SU(3)$ octet, ${{\cal O}}$, and an antidecuplet, $P$, ${\cal T}$ can be written as
\begin{eqnarray}
{\cal T}^{a,bc} \  =\  P^{abc}  \ +\ {1\over\sqrt{6}}\left(\ {{\cal O}}_d^{c}\ \epsilon^{abd}\ +\ {{\cal O}}_d^{b}\ \epsilon^{acd}\
\right)
\ \ \ .
\label{eq:Tantidefinedsu3}
\end{eqnarray}
This is precisely the interpolator for a degenerate ${\bf 8}\oplus\overline{\bf 10}$ which we constructed in
the diquark picture and is given in Eq.~(\ref{eq:pentaquarkinterp}).
Neglecting the kinetic term, the free Lagrangian is
\begin{eqnarray}
{\cal L}_{\,\uparrow} \ &=& \ -\ M_{{\bf 1}{\cal T}}^2\   {\cal T}^\dagger_{a,bc}\  {\cal T}^{a,bc} \ \ .
\label{eq:antilaghalfonesu3}
\end{eqnarray}
Clearly ${\cal O}$ and $P$ are degenerate.
As the octet and the antidecuplet are assumed to be $\textstyle{1\over2}^+$, explicit breaking will
induce mixing. If we turn on the quark mass matrix we must account for the operators
\begin{eqnarray}
{\cal L}^{M_q}_{\,\uparrow} & = &\ -\ {\alpha_{\cal T}}\ {\cal T}^\dagger_{a,bc}\, {\cal X}^a_{d}\,  {\cal T}^{d,bc}
\ -\ {\beta_{\cal T}}\ {\cal T}^\dagger_{a,bc}\, {\cal X}^b_{d}\,  {\cal T}^{a,dc}
\ -\ {\gamma_{\cal T}}\ {\cal T}^\dagger_{a,bc}\, {\cal T}^{a,bc}\, {\cal X}_d^{d} \ .
\label{eq:antiTlagsu3breaking}
\end{eqnarray}
Notice that the operator with coefficient ${\gamma_{\cal T}}$, while $O(M_q)$, is not $SU(3)$
violating and can therefore be absorbed into a redefinition of
$M_{{\bf 1}{\cal T}}^2$. We will nevertheless keep it for reasons of comparison.
The symmetry breaking operators induce mixing
between $N_{\tenb}$ and $N_{\cal O}$ and between $\Sigma_{\tenb}$ and
$\Sigma_{\cal O}$. Defining the mixed mass eigenstates $N_{1,2}$ and $\Sigma_{1,2}$ as
\begin{eqnarray}
N_{\cal O}\ &=&\ \sin\theta_N\ N_1\ +\ \cos\theta_N\ N_2 
\quad\,\,\, , \quad\,\,\,  
\Sigma_{\cal O}\ =\ \sin\theta_\Sigma\ \Sigma_1\ +\ \cos\theta_\Sigma\ \Sigma_2 
\nn \\
N_{\tenb}\ &=&\ -\sin\theta_N\ N_2\ +\ \cos\theta_N\ N_1 
\quad , \quad 
\Sigma_{\tenb}\ =\ -\sin\theta_\Sigma\ \Sigma_2\ +\ \cos\theta_\Sigma\ \Sigma_1 \ ,
\label{eq:antiTmixingpattern}
\end{eqnarray}
we then find off-diagonal operators of the form
\begin{eqnarray}
\textstyle{1\over{12}}\ (\ 2\alpha_{\cal T}\ -\ \beta_{\cal T}\ )\ ( m - m_s )
\left(\ 2\sqrt{2}\cos2\theta_\Gamma-\sin2\theta_\Gamma\ \right)\ {\Gamma^\dagger}_1 {\Gamma}^{}_2 \ +\ h.c.
\label{eq:antiTmixingoffdiag}
\end{eqnarray}
where $\Gamma\equiv N,\Sigma$. The mass-squared matrix is diagonalized by choosing $\sin\theta_\Gamma=-\sqrt{\textstyle{2\over3}}$
and $\cos\theta_\Gamma=\sqrt{1\over3}$, which corresponds to ideal mixing. 
Setting $\vev{c}={\bf 1}$, we then find the masses to be
\begin{eqnarray}
M_{N_1}^2\ &=& \ M_{{\bf 1}{\cal T}}^2\ +\ \alpha_{\cal T} \ m\ + \ \beta_{\cal T} \ m_s
\ + \ \gamma_{\cal T} \ (2m+m_s)\ , \nn \\
M_{N_2}^2\ &=& \ M_{{\bf 1}{\cal T}}^2\ +\ \alpha_{\cal T} \ m_s\  + \ \textstyle{1\over 2}\ \beta_{\cal T} \ (m+m_s)
\ + \ \gamma_{\cal T} \ (2m+m_s)\ , \nn \\
M_{\Sigma_1}^2\ &=& \ M_{{\bf 1}{\cal T}}^2\ +\ \alpha_{\cal T} \ m_s\ + \ \beta_{\cal T} \ m
\ + \ \gamma_{\cal T} \ (2m+m_s)\ , \nn \\
M_{\Sigma_2}^2\ =\ M_{\Lambda_{\cal O}}^2\ &=& \ M_{{\bf 1}{\cal T}}^2\ +\ \alpha_{\cal T} \ m\ 
+ \ \textstyle{1\over 2}\ \beta_{\cal T} \ (m+m_s) \ + \ \gamma_{\cal T} \ (2m+m_s)\ , \nn \\
M_{\Xi_{\tenb}}^2\ =\ M_{\Xi_{\cal O}}^2\ &=& \ M_{{\bf 1}{\cal T}}^2\ +\ \alpha_{\cal T} \ m\ + \ \beta_{\cal T} \ m
\ + \ \gamma_{\cal T} \ (2m+m_s)\ , \nn \\
M_{\Theta^{+}}^2\  &=& \ M_{{\bf 1}{\cal T}}^2\ +\ \alpha_{\cal T} \ m_s\ + \ \beta_{\cal T} \ m_s
\ + \ \gamma_{\cal T} \ (2m+m_s) \ .
\label{eq:antiTmixingmasses}
\end{eqnarray}
One finds, for instance, the mass-squared relations
\begin{eqnarray}
M_{N_1}^2\ + \ M_{\Sigma_1}^2\ &=& M_{N_2}^2\ +\ M_{\Sigma_2}^2\ = \ M_{\Xi_{\tenb}}^2\ +\ M_{\Theta^{+}}^2 \ .
\label{eq:antiTmixingmassesrelations}
\end{eqnarray}
It is easy to check that this mass-squared spectrum is equivalent to that found in Ref.~\cite{Jaffe:2003sg} by Jaffe and 
Wilczek~\footnote{We can choose our arbitrary constant $\gamma_{\cal T}=-\beta_{\cal T}$ so that
$M_{N_1}^2$ does not depend on the strange quark mass. Setting $m=0$ and making the identification
$M_0=M_{{\bf 1}{\cal T}}=\alpha_{\cal T}$ and $\alpha=-(\beta_{\cal T}/2\alpha_{\cal T}+1)m_s$
with $M_0$ and $\alpha$ the J-W parameters, one immediately recovers their results,
to $O(M_q)$.}. For discussion of phenomenology we refer the reader to Ref.~\cite{Jaffe:2003sg} and to Ref.~\cite{tom}.
It is worth recalling our input: we have assumed an octet and an antidecuplet in the irreducible
$(\overline{{\bf 3}},\overline{{\bf 6}})$ representation of $SU(3)\otimes SU(3)$ with purely-octet
symmetry breaking. 

\subsubsection{Axial Couplings in the Chiral Limit}

\noindent The main point of this paper is to observe that the chiral
multiplet structure which reproduces the ideally-mixed J-W model mass spectrum constrains 
the axial couplings as well.  We will now consider the
axial couplings in the chiral limit.  Now the left-handed current
transforms as an octet under $SU(3)_L$, a $(\bf{1},\bf{8})$, and the
right-handed current transforms as an octet under $SU(3)_R$, a
$(\bf{8},\bf{1})$.  We introduce $T^L_\alpha$ and $T^R_\alpha$, that
transform as
\begin{eqnarray}
T^L_\alpha\ \rightarrow \ L\  T^L_\alpha \ L^\dagger \qquad\qquad
T^R_\alpha\ \rightarrow \ R\  T^R_\alpha \ R^\dagger \ \ ,
\end{eqnarray}
under $SU(3)_R\otimes SU(3)_L$. 
The vector and axial-vector current matrix elements of the $(\overline{{\bf 3}},\overline{{\bf 6}})$ 
are reproduced by the effective currents
\begin{eqnarray}
T^{\,\uparrow}_\alpha 
& = & \  {\cal T}^\dagger_{a,bc} \left( T_\alpha \right)_d^{a}\  {\cal T}^{d,bc}
\ +\ 2\  {\cal T}^\dagger_{a,bc} \left( T_\alpha \right)_d^{b}\  {\cal T}^{a,dc}
\ \ \ , \nn \\
X^{\,\uparrow}_\alpha 
& = & \  {\cal T}^\dagger_{a,bc} \left( T_\alpha \right)_d^{a}\  {\cal T}^{d,bc}
\ -\ 2\  {\cal T}^\dagger_{a,bc} \left( T_\alpha \right)_d^{b}\  {\cal T}^{a,dc}
\ \ \ ,
\label{eq:antiTcurrentdefsu3}
\end{eqnarray}
respectively~\footnote{The coefficients of the operators
are obtained by taking matrix elements of the commutator of Eq.~(\ref{eq:basicchirasymmsu3}) between
various pentaquark states using Eq.~(\ref{eq:antiTcurrentdefsu3}).}.
Matching to the $\chi$PT axial-vector current of Eq.~(\ref{eq:chiptcurrentsu3pentas}) one directly
finds
\begin{eqnarray}
&&{\cal D}_{\cal O}\ =\ 1
 \ ,\  
{\cal F}_{\cal O}\ =\ -\textstyle{2\over 3}
 \ ,\  
{\cal H}_P\ =\ 1
 \ ,\  
{\cal C}_{P{\cal O}}\ =\ -2\textstyle{\sqrt{2\over 3}}
 \ ,\  \nn \\
&& \qquad\qquad {\cal C}_{PB}\ =\ {\cal D}_{{\cal O}B}\ =\ {\cal F}_{{\cal O}B}\ =\ 0
\label{eq:antiTaxialssu3}
\end{eqnarray}
in the chiral limit. The vanishing axial transitions from the pentaquarks to the baryons
is a simple consequence of LEMMA 2 as, by construction, the pentaquarks
and the baryons are in different chiral representations.
The only way to get ${\cal C}_{PB},{\cal D}_{{\cal O}B},{\cal F}_{{\cal O}B}\neq 0$ 
in the chiral limit without requiring pentaquark-baryon degeneracy 
is to place the pentaquarks and baryons in a reducible representation with 
pentaquark octet-antidecuplet mass-squared splitting of order ${{M}^{2}_{\bf{3}\bf{3}}}\sim \Lambda^2_{\scriptstyle QCD}$.
There is no natural way to maintain octet-antidecuplet degeneracy with non-vanishing couplings to the
ground-state baryons.

\subsubsection{Quark-Mass Corrections to the Axial Couplings}

\noindent Away from the chiral limit, mixing will shift the axial couplings within the
pentaquark chiral multiplet. Assuming that the ground-state
baryons are in a chiral multiplet that mixes with the pentaquarks at $O(M_q)$, then on general grounds one
expects
\begin{eqnarray}
{\cal C}_{PB}\ \ \mapright{M_q\neq 0} \ \bar{d}\ \ {{M^2_K}\over{\Lambda^2_\chi}}
\label{eq:cpbmqnotzero}
\end{eqnarray}
away from the chiral limit, where $\bar{d}$ is a dimensionless coupling of order one, $M_K=494~{\rm MeV}$ is the kaon mass
and $\Lambda_\chi\equiv 4\pi F_\pi$. (Notice that $\bar{d}$ contains a chiral logarithm as
one-loop effects contribute to the axial currents at $O(M_q)$ in the chiral expansion.) The total width
of $\Theta^+$ can be expressed as~\cite{Ko:2003xx,Mehen:2004dy,Mohta:2004yf}
\begin{eqnarray}
\Gamma(\Theta^+)\ =\ (146~{\rm MeV})\ {\cal C}_{PB}^2\  \mapright{M_q\neq 0}\ (5~{\rm MeV})\ \bar{d}^{\,2}  \ .
\label{eq:thetapluswidth}
\end{eqnarray}
where we have used Eq.~(\ref{eq:cpbmqnotzero}). Therefore, we find that in the natural J-W
scenario the $\Theta^+$ width is expected to be of order a few MeV. 

\section{Adler-Weisberger Sum Rules}
\label{sec:AW}

\noindent To get a better sense of what the predictions for the axial
couplings mean we will obtain the same results from a different
perspective.  Assume that the scattering amplitude for a pion
scattering on a hadron target with isospin-one in the t-channel falls
off sufficiently rapidly asymptotically to justify an unsubtracted
dispersion relation for the amplitude~\cite{AW,AW2}. The amplitude at
threshold simply measures the isospin of the target as guaranteed by
chiral symmetry low-energy theorems, while the dispersion integral can
be expressed as an integral over the total cross-section. Neglecting
the continuum and saturating the Adler-Weisberger sum rules for
elastic pion scattering on $p_{\cal O}$, $\Sigma^-_{\cal O}$,
$\Xi^-_{\cal O}$ and $N^+_{\tenb}$ with only those states within the
octet, ${\cal O}$, and the antidecuplet, $P$, yields
\begin{eqnarray}
1\ &=&\ ({\cal D}_{\cal O}\ +\ {\cal F}_{\cal O})^2\ +\ \textstyle{1\over 3}\ {\cal C}^2_{P{\cal O}} \ , \nn \\
2\ &=&\ \textstyle{2\over 3}\ {\cal D}^2_{\cal O}\ +\ 2\ {\cal F}^2_{\cal O}\ 
+\ \textstyle{1\over 6}\ {\cal C}^2_{P{\cal O}} \ , \nn \\
-1\ &=&\ -({\cal D}_{\cal O}\ -\ {\cal F}_{\cal O})^2\ +\ \textstyle{2\over 3}\ {\cal C}^2_{P{\cal O}} \ , \nn \\
1\ &=&\ \textstyle{1\over 9}\ {\cal H}_P\ +\ \textstyle{1\over 3}\ {\cal C}^2_{P{\cal O}} \ ,
\label{eq:pentaAW}
\end{eqnarray}
where the left-hand side measures twice the isospin of the target.
One readily checks that these Adler-Weisberger sum rules have two solutions:
\begin{eqnarray}
&&|{\cal D}_{\cal O}|\ =\ 1
 \ ,\  
|{\cal F}_{\cal O}|\ =\ \textstyle{2\over 3}
 \ ,\  
|{\cal H}_P|\ =\ 1
 \ ,\  
|{\cal C}_{P{\cal O}}|\ =\ 2\textstyle{\sqrt{2\over 3}} \ ,
\label{eq:AWsolution1}
\end{eqnarray}
and
\begin{eqnarray}
&&|{\cal D}_{\cal O}|\ =\ 0
 \ ,\  
|{\cal F}_{\cal O}|\ =\ 1
 \ ,\  
|{\cal H}_P|\ =\ 3
 \ ,\  
|{\cal C}_{P{\cal O}}|\ =\ 0 \ .
\label{eq:AWsolution2}
\end{eqnarray}
The first solution corresponds ---up to phases that are further
constrained by inelastic Adler-Weisberger sum rules--- to placing
${\cal O}$ and $P$ in a $(\overline{{\bf 3}},\overline{{\bf 6}})$
representation.  One may easily verify that the second solution
corresponds to the only other possibility: putting ${\cal O}$ in a
$({\bf 1},{\bf 8})$ and $P$ in a $({\bf 1},\overline{{\bf 10}})$.  Of
course the second solution does not require octet-antidecuplet
degeneracy.  Notice that the coupling between the octet and
antidecuplet pentaquarks in the second solution vanishes. This is once
again due to LEMMA 2.

\section{Discussion}
\label{sec:conc}

\noindent The diquark model of pentaquarks of Jaffe and
Wilczek~\cite{Jaffe:2003sg} has a remarkably simple and natural
interpretation in terms of the $SU(3)_R\otimes SU(3)_L$ chiral
symmetry of QCD.  By placing an ${\bf 8}$ and a ${\bf\overline{10}}$
of flavor-$SU(3)$ in a $(\overline{{\bf 3}},\overline{{\bf 6}})$ of
$SU(3)_R\otimes SU(3)_L$ one enforces ${\bf 8}-{\bf\overline{10}}$
degeneracy without resorting to quark-model assumptions. When the
quark masses are turned on one finds precisely the ideally-mixed mass
spectrum of Ref.~\cite{Jaffe:2003sg}.  Moreover, the chiral-multiplet
structure completely determines the axial-vector transitions of the
pentaquarks.  In particular, one finds that the pentaquarks do not
communicate with the baryons by Goldstone-boson exchange in the chiral
limit.  Hence, the pentaquark widths scale as the quark mass matrix
squared and are generically small, of order a few MeV.  It should be
noted that analogous arguments hold for tetraquark and pentaquark
nonets built out of diquarks, which naturally fall into
$(\bf{\overline{3}},\bf{3})$ and $(\bf{3},\bf{\overline{3}})$
irreducible representations of $SU(3)_R\otimes SU(3)_L$~\footnote{
This prediction is strongly at odds with the tetraquark interpretation of the controversial broad scalar
$f_0(600)$.}.

It may well be the case that the true chiral multiplets in which the
pentaquarks find themselves are quite different than what is suggested
by the diquark picture. However, as long as the baryons and the
pentaquarks are not in the same chiral multiplet, the estimate of the
width given in Eq.~(\ref{eq:thetapluswidth}) continues to hold. For
instance, the pentaquark octet may be absent, as is the case in the
chiral-soliton
model~\cite{Manohar:1984ys,Chemtob:1985ar,Prasza,Diakonov}.  The
antidecuplet $P$ may then find itself, for instance, in the $({\bf
1},\overline{{\bf 10}})$ of $SU(3)_R\otimes SU(3)_L$ which again
experiences no axial transitions to the baryons in the chiral
limit. However, as no degeneracies among flavor multiplets are
expected in the chiral-soliton model, there is no compelling rationale
to conclude anything simple about its chiral multiplet structure.

\acknowledgments

\noindent I would like to thank Bob Jaffe for a discussion which motivated this work and
for providing useful comments on the manuscript. This work was supported in part 
by the National Science Foundation under grant No. PHY-0400231 and by DOE contract DE-AC05-84ER40150,
under which the Southeastern Universities Research Association (SURA)
operates the Thomas Jefferson National Accelerator Facility. 


\end{document}